# Evidences for Skeletal Structures in the Ocean from the Images Analyzed by Multilevel Dynamical Contrasting Method

V. A. Rantsev-Kartinov

INF RRC "Kurchatov Institute", Moscow, 123182, Russia
rank@nfi.kiae.ru

**Abstract.**

An analysis of databases of photographic images of ocean's surface, taken from various altitudes and for various types of rough ocean surface, revealed the presence of an ocean's skeletal structures (OSS), which exhibit a tendency toward self-similarity of structuring at various length scales (i.e., within various «generations»). The topology of OSS appears to be identical to that of skeletal structures (SS) which have been formerly found in a wide range of length scales, media and for various phenomena. The typical OSS consists of separate identical blocks which are linked together to form a network. Two types of such blocks are found: (i) coaxial tubular (CT) structures with internal radial bonds, and (ii) cartwheel-like (CW) structures, located either on an axle or in the edge of CT block. The OSSs differ from the formerly found SSs only by the fact that OSS, in their interior, are filled in with closely packed OSSs of a smaller size (i.e. OSSs of former «generations»). We specially discuss the phenomenon of skeletal blocks in the form of vertically/horizontally oriented floating cylinders (VFC/HFC). The size of these blocks is shown to grow with increasing rough water.

## 1. Introduction.

The author's studies of skeletal structures have started from analyzing the photographic images of the plasma with the help of the method of multilevel dynamical contrasting (MMDC) [1a,b]. This method is based on the variability of the computer-made maps of contrasting the image (the above variability is needed to avoid the artifacts resulted, sometimes, from contrasting the image). The MMDC allows (a) to resolve the fine structure of separate formations, hardly seen in the images, and reveal the almost imperceptible links between such; (b) reveal the OSS even at some depth under water surface. The MDC allows to significantly eliminate the masking background in the image and identify the correlation, if any, between the elements.

The skeletal structures (SSs) -- namely, tubules, "cartwheels" and their simple combinations, with certain properties (longevity of these structures, a trend toward self-similarity of their structuring, and toward their self-assembly) -- have been found in a very wide range of phenomena in laboratory, Earth atmosphere and space [1-3]. The formulation of hypotheses [1c,d,f,g] for the exceptional role, in the observed phenomena of a SS, of quantum bonds, which are presumably provided by a nanodust (presumably, carbon nanotubes or similar nanostructures), allowed to explain some properties of SSs. An extension of major hypothesis [1c,d,f,g] to a very wide range of length scales have covered [1e, 2a] as much as about 30 orders of magnitude.

In the present paper it is shown that similar skeletal structures are found in the ocean. We call them an ocean's skeletal structures (OSS). The OSSs differ from the formerly found SSs only by the fact that OSS, in their interior, are filled in with closely packed OSSs of a smaller size (i.e. OSSs of former «generations»).

## 2. Hypotheses for the formation of ocean skeletal structures (OSS).

According to [1e,h;2a,c], *tornado - a special dusty plasma in the atmosphere with the key role of nanodust.* This conjecture leads to seeking for the source and the nature of the structure-producing dust. The study of trajectories of tropical hurricanes for more than a century period of observations suggests the hypothesis, **I,** *for a direct link between the origin of severe weather phenomena (SWP) and the location of intense volcanic activity*.

This hypothesis naturally follows from [1e,h;2a,c] because the volcano may inject structure-producing dust to atmosphere. As far the oceanic surface amounts to 2/3 of the Earth surface, a question naturally arises about the probable implications of the presence of such a dust, and of SSs, in the upper layers of the ocean and in the lower atmosphere. Therefore, the established approaches to dynamics in these regions may be substantially influenced by this factor, especially if we will not ignore the impact of electromagnetic factors on the weather in these regions and allow for the possibility of self-assembly of SS in the presence of an electric field. This gives the hypothesis, **II,** *for the role of electric field of the Earth, and of atmospheric electricity, in forming the SS which may invoke the SWP.*

The SS possesses certain mechanical strength and exceptional chemical, physical, and electromagnetic properties. The presence of such SS in the sea water which possesses electrolytic properties and contains mineral and biological components, may give many interesting implications. This suggests the hypothesis, **III,** *for initiating, by the SS immersed in the sea, of a number of phenomena, namely osmosis, division of electric charges, formation of double layers etc.*

The presence of carbon nanotubes of small diameter in SS assumes that they are not fully filled in with sea water, because of the presence of water clusters and particulate of phytoplankton. This gives the hypothesis, **IV,** *for the buoyancy of SS in the sea water.*

This idea is supported by the adsorption of air bubbles in the water by the SS to give a partial flotation of SS. In the sea water, various substances in different phase states of matter are in touch. This may substantially change the formerly suggested scenarios for SS formation in other cases. This suggests the hypothesis, **V**, *for the possibility of the action of surface tension even on the blocks of SS which is immersed in the sea*. This phenomenon results in the aggregation of blocks deposited from the atmosphere on the ocean's surface, to form eventually an OSS. The skeletal structuring assumes that the individual straight and rather strong blocks may be joined flexibly, similarly to joints in a skeleton.

## 3. Evidences for SSO.

The identification of the OSS is attractive in a sense that this is feasible for everybody interested because, at preliminary stage, this does not require special diagnostic apparatus and skills, and may give results easily. This, however, requires anyway (i) having a high-resolution, low-contrast image and (ii) a priori knowledge of the structure under search, (iii) a capability of being a good observant.

We start with evidences for OSS in the case of cartwheel (CW) structures which often appear at the ocean surface. An example, the ocean surface for the case of the force-one wind, is given in Fig. 1.

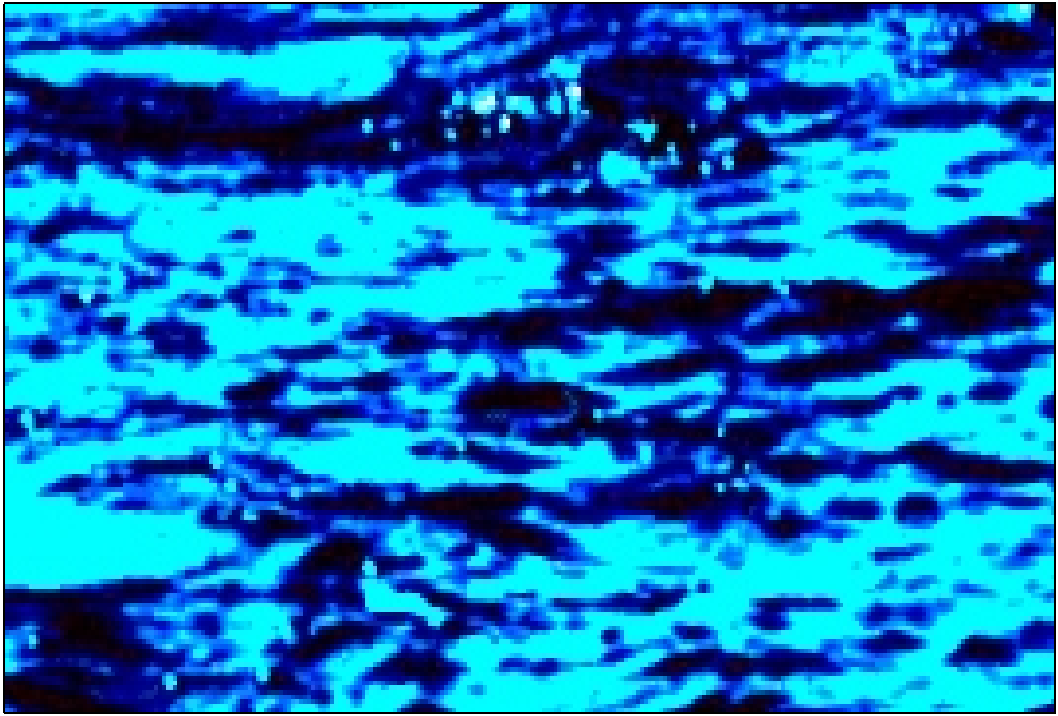

**Fig. 1**. An example of a cartwheel (CW) structure under the force-one wind (image's width is 4 m). The CW structure, seen in the center of the image, is distinct even without MDC processing. The outer diameter of CW is ~ 2 m. Cylindrical (tubular) blocks which compose this structure, are of 5-10 cm in diameter; diameter of the near-axis region is ~ 40 cm.

It is seen that the CT seems to be not a separate block but a part of an OSS whose elements may also be seen on the surface. This structure seems to be not a vortex because it fwas a steady-state one during the exposure time. The side surface and near-axis part of the structure are located a little bit above sea-level, that also differs from the case of a vortex. The major difference of CW from the vortex is the presence of distinct radial bonds/spokes between the axle and the rim, with the spoke being sometimes piercing the rim. These facts are in odds with hydrodynamic origin of such structures.

Now let's turn to evidences for straight coaxial - tubular (CT) blocks in OSSs. The CT blocks may be of various size, may have various structure of side surface (varying from smooth or waver-like to a braided or squirrel wheel-like form), may be floating on the surface with horizontal or vertical orientation of axis (i.e. being a vertically/horizontally oriented floating cylinders (VFC/HFC)), may be of a multi-layer structure and possess either radial bonds or a CW in the edge. A slightly sunk VFC with CW in its edge, which may be seen in the rough water, looks like a structure seen in Fig. 1. Sometimes, a part of the VFC of larger diameter is located fully under water or visible partly, if located slightly above the sea level. In such a case, the inner part of smaller diameter may look like a vertically oriented fishing float. An example of such a VFC is given in Fig. 2.

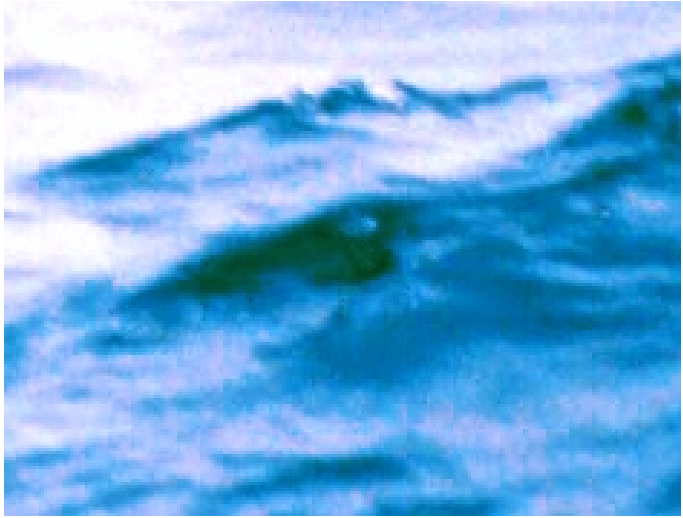

**Fig. 2.** An example of vertically oriented floating cylinder (VFC) (image's width 1 m). In the center the image, one may see a structure ~40 cm above the sea level, with almost orthogonal branches. There is a dip in the center of the structure. Within the dip, there is a vertically oriented rod, like a fishing «float», of 6 cm diameter. The float has a thinner rod of diameter ~ 1 cm, which juts out.

Now we turn to what happens with OSS and its blocks under sea storm conditions. An analysis of photographic images of the sea surface, taken by the flying laboratory in stormy regions, lead us to a conclusion that a strong wind strips the OSS, with the size of the stripped blocks being increased with the increasing wind. To demonstrate the evolution of OSS blocks stripped by the storm, let us look at the above-mentioned type of VFC. The fragment of the images of sea surface, taken during the Hurricane Belle from 500 feet altitude (NOAA collection [4]), is given in Fig. 3.

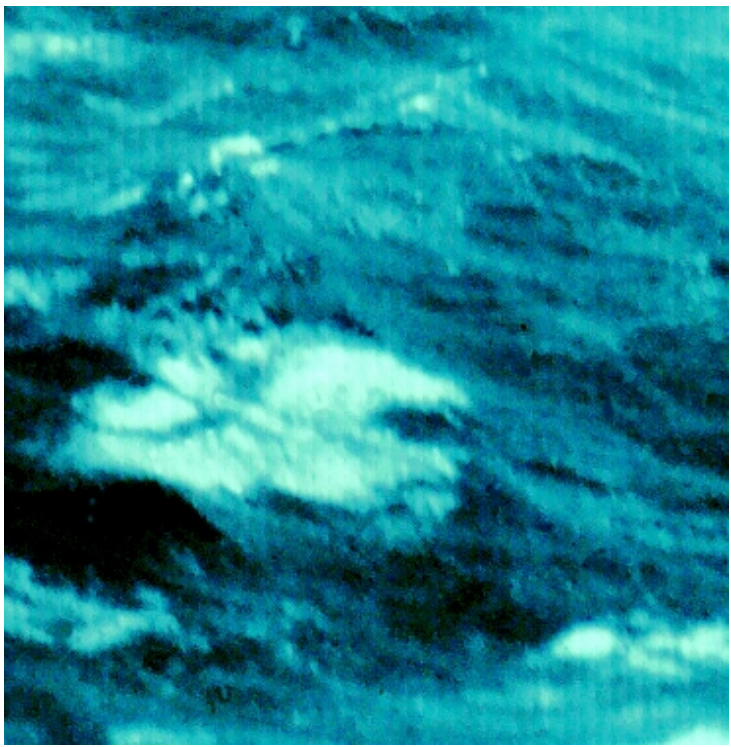

**Fig. 3.** A fragment of the image of sea surface, taken during the Hurricane Belle from 500 feet altitude (NOAA collection [4]). Image's width is ~15 m. One may see, in the front of the image, almost a half of a vertically oriented floating cylinder (VFC), of diameter ~ 10 m. Diameter of central axial tube is ~ 2 m, diameter of dark rings in the edge of VFC is ~1 m. The VFC is ~ 1 m above the sea's local level.

An example of the image of a stormy ocean surface was found by the author in the available databases, which -- when MDC processed -- shows a row of four identical VFC, whose edges sustain the parallel orientation under a strong storm, thus suggesting the presence of a strong link between these VFC. An interesting fragment from similar data [5] is given in Fig. 4. The original image was taken in the center of the Hurricane Caroline. The image demonstrates an increase of diameter of CW structure, under hurricane conditions, up to ~10 meters, i.e. larger by a factor of 6, as compared to that at conditions of the relatively quiet sea.

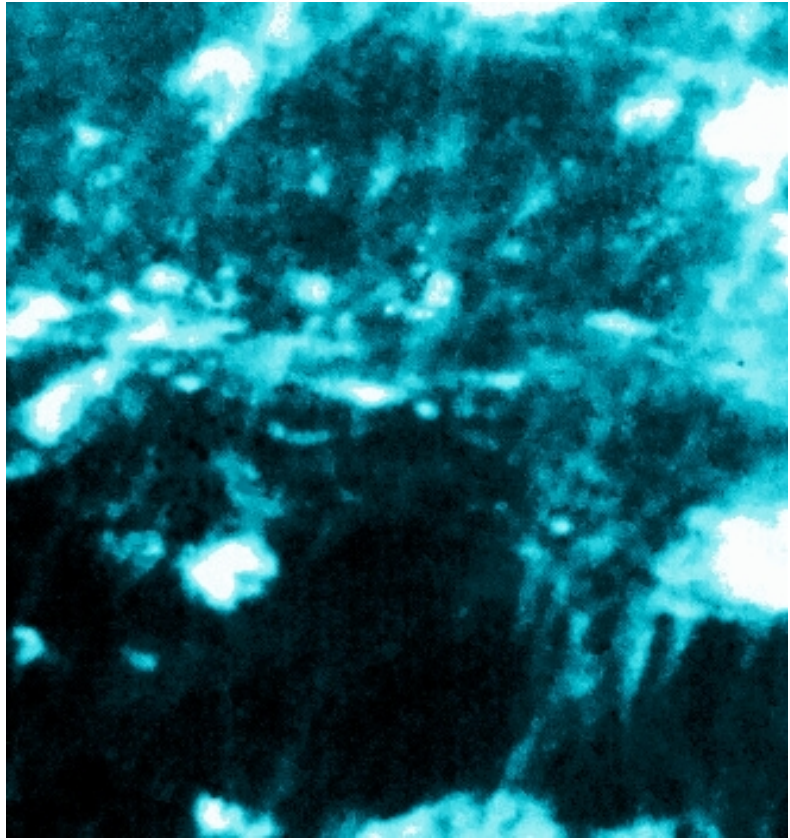

**Fig. 4.** A fragment of the image of sea surface, taken in the center of the Hurricane Caroline (NOAA collection [5]). Image's width is ~ 15 m. There is a distinct cartwheel (CW) structure of diameter ~ 10 m in the upper part of the image. This CW seems to be a second storey with respect to a dark coaxial tubular (CT) structure which is ahead of CW by few tens of meters. Between CT and CW, there is a boundary between two waves directed to the observer to the left.

The crest often appears to be a horizontally oriented floating cylinder (HFC), with the water wave rolling over HFC's side surface, i.e. transversely to HFC's axis. In the case of a collision of two mutually orthogonal HFCs, a disruption of the envelope of one of them reveals the internal structure of HFC. An analysis of such structures suggests that the large scale cylinders, observed in the waves in a stormy ocean, seem to be a multi-layer telescopic tubes with internal radial bonds. It appears that the ratio of diameters for neighboring layers amounts to a factor of 2. Such a phenomenon is illustrated with Fig. 5.

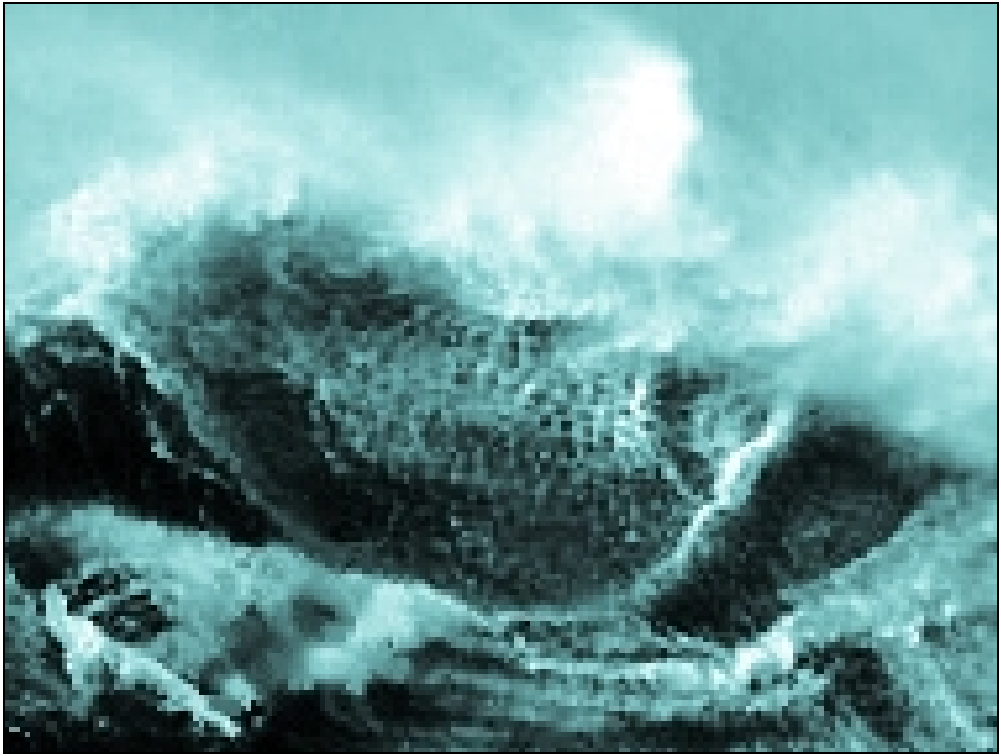

Fig. 5. The image of a horizontally oriented floating cylinder (HFC) whose envelope is disrupted. Image's width is ~ 6 m. At the image's rear, one may see a disruption of a cylinder, as a cut in cylinder's horizontal direction. Also some details of internal structure of HFC are seen, namely: the remains of radial spokes between internal and outer shells; a net-like structure of internal cylinder. Diameter of the entire cylinder is ~ 3 m, and those of internal nested shells are ~ 1.5 m, 0.75 m, and 0.3 m, respectively (i.e., the ratio of neighboring shells' diameters is a factor of ~ 2).

## 4. Conclusion and comments.

The results of the present research allow to suggest an existence of the ocean skeletal structures (OSS). The separately floating blocks of OSS, which are found in the available databases, tend to form a structure determined by the local level of sea waves. The time of formation of a stable and long-livid structure seems to be finite and strongly dependent on the weather conditions and local physics parameters. This allows to explain the longevity (seen for a day) of the traces of sea ships, up to half a thousand of kilometers in length - as it is seen from observations of the ocean, during the calm time, from space satellites. Indeed, as far as the reflectivity of ocean surface may depend on the hypothetical presence of OSSs and their structure, the ship's water screw may destroy the homogeneity of the established structuring on the ocean's surface, and some time is needed to come back to the undisturbed state. It looks like the peculiar features of the structure in a certain region are determined by the history of its formation.

The author is deeply grateful to A.B. Kukushkin for a decade long collaboration. Special thanks to V.I. Kogan for invariable support and interest to a research of skeletal structures.

# REFERENCES

[1] A.B. Kukushkin, V.A. Rantsev-Kartinov, a) Laser and Particle Beams, **16**, 445,(1998); b) Rev.Sci.Instrum., **70**, 1387,( 1999); c) Proc. 17-th IAEA Fusion Energy Conference, Yokohama, Japan, **3,** 1131, (1998); d) Proc. 26-th EPS PPCF, Maastricht, Netherlands, 873, (1999); e) Phys. Lett. A, **306**, 175 (2002); f) Current Trends in International Fusion Research: Review and Assessment (Proc. 3$^{rd}$ Symposium, Washington D.C., March 1999), Ed. E. Panarella, NRC Research Press, Ottawa, Canada, 121, (2002); g) Preprint of Kurchatov Institute, IAE 6111/6, Moscow, October (1998); h) "Advances in Plasma Phys. Research", (Ed. F. Gerard, Nova Science Publishers, New York), **2**, 1, (2002);.

[2] A.B. Kukushkin, V.A. Rantsev-Kartinov, (a) Sicence in Russia, **1**, 42, (2004); (b) Microsystem technology (in Russian), **3**, 22 (2002); (c) Mat. IV Russian Seminar «Modern methods of plasma diagnostics and their application to substance probing and environmental control», MEPhI, Moscow, 2003, p. 151.

[3] B.N. Kolbasov, A.B. Kukushkin, V.A. Rantsev-Kartinov, P.V. Romanov, Phys. Lett. A, **269**, 363, (2000); Ibid., **291,** 447, (2001); Plasma Devices & Operations, **8** , 257, (2001).

[4] http://www.photolib.noaa.gov/flight/images/big/fly00164.jpg

[5] http://www.photolib.noaa.gov/flight/images/big/fly00166.jpg